\documentclass[aps,prd,twocolumn,groupedaddress,showpacs,nofootinbib,floatfix]{revtex4}
\usepackage{epsfig}
\usepackage{subfigure}
\usepackage{dcolumn}
\usepackage{bm}

\begin{document}
\preprint{NIKHEF/2003-017}
\preprint{BI-TP 2003/37}
\title{Pion structure from improved lattice QCD:\\
        form factor and charge radius at low masses}
\author{J. van der Heide}
\email{r86@nikhef.nl}
\affiliation{National Institute for Nuclear Physics and High-Energy
  Physics (NIKHEF), 1009 DB Amsterdam, The Netherlands}
\author{J.H. Koch}
\email{justus@nikhef.nl}
\affiliation{NIKHEF and Institute for Theoretical Physics, University of
  Amsterdam, Valckenierstraat 65, 1018 XE Amsterdam, The Netherlands}
\author{E. Laermann}
\email{edwin@physik.uni-bielefeld.de}
\affiliation{Fakult\"at f\"ur Physik, Universit\"at Bielefeld, D-33615
  Bielefeld, Germany}
\begin{abstract}
The charge form factor of the pion is calculated in 
lattice QCD. The non-pertur\-ba\-tive\-ly improved
Sheikholeslami-Wohlert action is used together 
with the $\mathcal{O}(a)$ improved vector 
current. Other choices for the current are examined. 
The form factor is extracted for pion 
masses from 970 MeV down to 360 MeV and for momentum 
transfers $Q^2 \leq 2 \;\mathrm{GeV}^2$. 
The mean square charge radius is extracted, 
compared to previous determinations and its 
extrapolation to lower masses discussed.\\
\end{abstract}

\date{\today}
\maketitle


\section{Introduction}
QCD without doubt is the correct microscopic theory 
describing all strong interactions, a fact that has 
been established mainly by impressive agreement between 
theory and experiment in the perturbative sector. 
Comparatively few results were obtained in 
non-perturbative QCD, which deals with physics on the 
scale of $\Lambda_{QCD}$ or the size of a hadron. 
It is therefore an obvious challenge to derive the 
internal structure of a hadron from first principles, 
entirely within QCD \cite{Barad:1984px}. 
Next to the nucleon, the pion is 
an obvious candidate for such an attempt. Its global 
features, like charge, spin and isospin, represent no 
challenge and are trivially included in any model. 
Specific features that are actually testing details of 
our theoretical understanding are observables like the 
pion form factor or its polarizability. In this paper, 
we report on an extensive study of the pion form factor 
based on lattice QCD. First results were already 
reported in~\cite{vanderHeide:2003ip,vanderHeide:2003vz}.

At first glance, the pion looks like a manageable two-body 
system and there have been many descriptions of the pion 
based on effective models or QCD inspired approaches. 
One feature all these attempts 
share is that confinement, the most striking feature of QCD, 
is - in one way or another -  put in by hand. This is of 
course an unwanted step when one sets out to calculate the 
pion form factor or its mean square charge radius, which 
reflect the form and size of QCD confinement. Here one 
obviously wants to proceed from first principles, from the 
fundamental QCD Lagrangian itself. 

Several papers have already dealt with aspects of the pion 
structure in lattice QCD. An often considered quantity is 
the 'Bethe-Salpeter amplitude', the relative quark-antiquark 
wave function, extracted from 2-point functions
~\cite{Chu:1991ps,Hecht:1992uq,Gupta:1993vp,Laermann:2001vg}. 
Another approach has been based on gauge invariant density-density 
correlations \cite{Barad:1984px,Chu:1991ps,Wilcox:1986ge},
most recently by 
Alexandrou \textit{et al.}~\cite{Alexandrou:2002nn}. 
Two groups have already calculated the pion 
charge form factor, proceeding, as in the present work, in the 
quenched approximation. The pioneering work was done by 
Martinelli \textit{et al.}~\cite{Martinelli:1988bh}, 
followed by more detailed calculations of 
Draper \textit{et al.}~\cite{Draper:1989bp}. One of 
the findings of the latter work was that the pion form 
factor could be described quite well by a monopole 
form, as suggested by vector meson dominance. As shown 
in~\cite{vanderHeide:2003ip}, the range parameter is in fact 
very close to the $\rho$-mass obtained for the same action.

Lattice calculations, although starting from first principles, 
are not free of approximations. The most obvious one is the 
use of the lattice itself, necessarily resulting in 
discretization errors. These errors can be reduced by the use 
of improved lattice QCD actions and the concomitant improved 
observables. In the work reported here, we extend the 
previous work by working in ${\cal O}(a)$ improved 
lattice QCD, which guarantees that errors in the matrix 
elements we extract are only of order ${\cal O}(a^2)$. 
In order to emphasize the importance of consistently using 
both improved action and observables, we discuss the form 
of the vector current operator at some length and give numerical 
examples for results one obtains with the current operators 
used in other work. 

In addition to the step from a discrete lattice to the 
physical continuum, one also has to extrapolate the lattice 
results in the pion mass. As is known, lattice calculations 
yield results for pions much heavier than the physical pion. 
The previous form factor calculations 
in~\cite{Martinelli:1988bh,Draper:1989bp} were for pions 
on the order of 1 GeV. Another improvement step which we 
undertake in this paper is to extend our calculations of 
the electric form factor down to pion masses of 360 MeV. 
For the mean square charge radius of the pion which we 
extract from the form factor, we then study the extrapolation 
to lower masses. 

It is instructive to compare the mean square radii obtained 
from the Bethe-Salpeter amplitudes and from the form factor. 
We find considerable differences which, as suggested 
earlier~\cite{Gupta:1993vp} can be ascribed to the effect 
of the gluon motion on the position of the center of mass.

The paper is organized as follows. In Ch. 2, we first 
describe the general features of our approach and the 
details of our lattice calculations. Results for the 
2- and 3-point function are described 
in Ch. 3 and 4, respectively. Our findings for the form factor 
and the mean square radius of the pion are elaborated 
on in Ch. 5. A summary of our work and conclusions are 
contained in Ch. 6.

\section{The general method}

\subsection{The observables}

To obtain the pion form factor from the lattice, one has to 
calculate two observables, 
the 2- and the 3-point Green's function for an 
interacting quark-antiquark pair with the quantum numbers of 
the pion. Improved techniques are used in order to 
remove $\mathcal{O}(a)$ discretization effects. We do this 
non-perturbatively for both the action and the electromagnetic 
vector current in the 3-point function. Fits to both 
observables were used to extract the desired information, 
such as the form factor and the pion mass.

\begin{figure}[h]
\includegraphics[width=75mm]{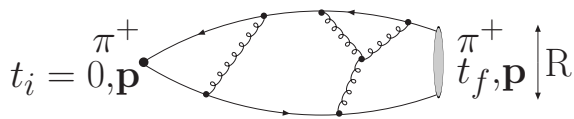}
\caption{\label{fig:2p}2-point function.}
\end{figure}

The 2-point function, shown schematically in Fig.~\ref{fig:2p},  
projected to momentum ${\bm p}$ is given by
\begin{equation}
G_{2,R}(t_f,\bm{p}) = \sum_{\bm{x}} \left<\phi_R(t_f, \bm{x})\; \phi ^{\dagger}
(0,\bm{0})\right> \; e^{i\; \bm{p}\cdot \bm{x}} \; .
\label{eq:2p}
\end{equation}
The operator $\phi^{\dagger}$ creates a quark-antiquark pair with the 
quantum numbers of the pion at the source at $(0,\bm{0})$, while $\phi (x)$
annihilates it at the sink. Since we will consider a $\pi ^+$, it is given by 
\begin{equation}
\phi^\dagger (x) = {\bar \psi}_u (x)\; \gamma_5\; \psi_d (x) \; .
\label{eq:pion_op}
\end{equation}
Below, all flavor, spin and $SU(3)$ color indices will be dropped. 
On the sink side, we use
an extended operator with an inter-quark distance $R$. 
This suppresses the contribution of excited states
to the 2-point function and facilitates the extraction 
of the pion mass.
In order to keep the calculation 
gauge invariant, quark and anti-quark at the sink are connected
by gauge links. To further enhance the contribution from the pion, the links 
in the extended pion operator have been fuzzed to 
better simulate the tube-like nature of the gluon cloud. 

\begin{figure}
\includegraphics[width=85mm]{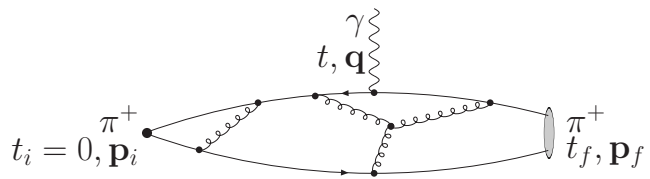}
\caption{\label{fig:3p} 3-point function.}
\end{figure}

The 3-point function, shown in Fig.~\ref{fig:3p}, again concerns a 
quark-antiquark pair with pion quantum numbers, propagating 
from $x_i$ to $x_f$; disconnected diagrams do not contribute 
~\cite{Barad:1984px,Draper:1989bp}.
At an intermediate time $t$ a photon is 
coupled to either one of the
charged quarks. 
This observable is obtained in momentum space by calculating 
\begin{eqnarray}
G_{3,\mu} (t_f, t; \bm{p}_f , \bm{p}_i) = \sum_{\bm{x}_f,\bm{x}} \left<\phi_R(x_f) \; j_{\mu} (x) \; \phi^{\dagger} (0)\right> \nonumber \\
\times \;\; e^{-i\; \bm{p}_f \cdot (\bm{x}_f - \bm{x}) \; - i \; \bm{p}_i \cdot \bm{x}} \;.
\label{eq:3p}
\end{eqnarray}

The parameter $R$ for the pion operator at the sink is 
now fixed to the value giving the best overlap in the
2-point function. As we will further discuss 
below, the choice of the current to which the quarks couple is important. 
The continuum or {\it local} current, 
\begin{equation}
j_{\mu}^L=\bar{\psi}(x)\gamma_{\mu}\psi(x)\; , 
\label{eq:lc}
\end{equation}
is not conserved on the lattice and needs 
renormalization by a factor $Z_V$, yielding the {\it renormalized
local} current $j_\mu^{RL}$. Using the Noether 
procedure, one can also construct a current which 
is conserved on the lattice~\cite{Karsten:1981wd}, 
\begin{eqnarray}
j^{C}_\mu={\bar\psi(x)}(1-\gamma_\mu)\,U_\mu(x)\,\psi(x+{\hat \mu})~~~~~~~~~~~~~ \nonumber\\
-{\bar\psi(x+{\hat\mu})}(1+\gamma_\mu)\,U^{\dagger}_\mu(x)\,\psi (x) 
\;.
\label{eq:cc}
\end{eqnarray}
This {\it conserved} current still requires $\mathcal{O}(a)$ discretization corrections 
for matrix elements away from the forward direction.

Using Symanzik's improvement program, one can identify appropriate 
operators, which,
when used together with the improved action, result in matrix elements that 
are free of  all $\mathcal{O}(a)$ discretization 
errors. For the vector current considered here, 
the resulting {\it improved} current is
\cite{Martinelli:1991ny,Luscher:1997jn,Guagnelli:1998db}
\begin{equation}
j^I_{\mu} = Z_{V} \, \{ j^{L}_\mu (x)  + a \, c_V \, \partial _\nu \, 
T_{\mu \nu} \} \;, 
\label{eq:ic}
\end{equation} 
with 
\begin{eqnarray}
T_{\mu \nu} & = & {\bar \psi}(x)\; i \; \sigma_{\mu \nu}  \; \psi (x) 
\ ,
\label{eq:renorm}\\
Z_V & = & Z_V^0 \: (1+ a\,b_V\,m_q) \; .\nonumber
\end{eqnarray}
It is conserved to $\mathcal{O}(a^2)$ and differs from the renormalized 
local current by a total divergence, which vanishes for forward matrix elements.
The bare quark mass is defined as
\begin{equation}
a\;m_q=\frac{1}{2}\left(\frac{1}{\kappa}-\frac{1}{\kappa_{c}}\right)\, ,
\end{equation}
where $\kappa_{c}$ is the kappa value in the chiral limit and $a$ the 
lattice spacing. The constants in $j^I_{\mu}$ are non-perturbatively 
determined from lattice simulations and as such completely remove 
the ${\mathcal{O}}(a)$ effects.

From current conservation it can be shown that the general Lorentz 
structure of the matrix element for the electromagnetic current of 
an on-shell pion is 
\begin{equation}
\left<\pi(\bm{p}_f)|j_\mu|\pi(\bm{p}_i)\right>_{cont}=(p_f+p_i)_\mu\;F(Q^2)\;,
\label{eq:lorentz_dec}
\end{equation}
where $F(Q^2)$  with $Q^2 = - (p_f - p_i)^2$ is the form factor we are interested in.
Connection with the continuum description is made by a proper normalization,
\begin{equation}
\left<\pi(\bm{p}_f)|j_\mu|\pi(\bm{p}_i)\right>_{latt.}=
\frac{\left<\pi(\bm{p}_f)|j_\mu|\pi(\bm{p}_i)\right>_{cont.}}{2\sqrt{E_f E_i}}\;,
\label{eq:normalisation}
\end{equation}
where $E_f$ and $E_i$ are the final and initial energies, respectively. 

In our calculations of the 3-point function 
we project on initial and final three momenta with the same length,
\begin{equation}
|\bm{p}_i| = |\bm{p}_f|\;,
\end{equation}
which implies that there is no energy transfer to the pion,
\begin{equation}
E_i = E_f\;. 
\end{equation}
The 4-momentum transfer to the pion, $Q^2=(\bm{p}_f-\bm{p}_i)^2$,
is then varied by changing the angle between the two momenta. Since we
will use the $\mu = 4$ component of the current, 
this choice has, among others, the numerical advantage that we have
\begin{equation}
\frac{E_f + E_i}{2{\sqrt{E_f E_i}}} = 1 
\label{eq:prefac}
\end{equation}
when extracting the form factor $F(Q^2)$.

\begin{figure}
\includegraphics[width=50mm]{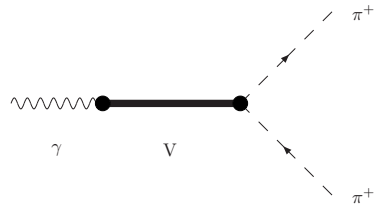}
\caption{\label{fig:vmd} The VMD model.}
\end{figure}

The Vector Meson Dominance (VMD) model has been 
quite successful in describing both experimental as well as early 
lattice data. This model is inspired by effective field theory and is 
schematically depicted in Fig.~\ref{fig:vmd}. 
Assuming universality, \textit{i.e.} $g_{\rho\pi\pi}=g_{\rho}$,
the form factor is given by the simple monopole form 
\begin{equation}
F(Q^2)=\left\{1+\frac{Q^2}{m_{V}^2}\right\}^{-1} \; .
\label{eq:vmd}
\end{equation}

\subsection{The lattice simulation}

We performed calculations in the quenched approximation on a $24^3\times32$ 
lattice. A set of 100 gluon configurations at a coupling of $\beta=6.0$ was 
generated. Thermalization was reached in 2500 sweeps, whereafter we 
obtained configurations at intervals of 500 sweeps. One sweep consists of a 
pseudo-heatbath step with FHKP updating in the \textit{SU}(2) subgroups, 
followed by four over-relaxation steps. 

We used the improved Sheikholeslami-Wohlert action~\cite{Sheikholeslami:1985ij}
with the non-perturbatively determined ~\cite{Luscher:1997ug}
value of $c_{SW}=1.769$. With this action, we computed propagators for five 
values of the hopping parameter corresponding to pion masses\footnote{For definiteness we have taken $a=0.105 \;fm$ 
from~\cite{Edwards:1998xf} to set the scale.} of 970, 780, 670, 540 and 360 
MeV, see Table 1. We imposed periodic boundary conditions except in the 
time direction where for the fermions anti-periodic boundary conditions were 
implemented. The values of the constants $c_V, b_V$ and $Z_V^0$ needed to
eliminate the ${\cal O}(a)$ effects and to renormalize the current
were taken from Bhattacharya \textit{et al.}~\cite{Bhattacharya:2000pn}.

\section{The 2-point function}
To extract physical information from the numerical data for the 2-point 
function, Eq.~\ref{eq:2p}, we use the following parametrization 
\begin{eqnarray}
G_{2,R} (t_f,\bm{p}) = \sum_{n=0}^1{\sqrt {Z_R^n(\bm{p})\: Z_0^n(\bm{p})}} 
\; e^{-E^n_{\bm{p}}\;\frac{N_\tau}{2}} \nonumber \\
\times \;\; \cosh \{E^n _{\bm{p}}\:(\frac{N_\tau}{2} - t_f)\} \; ,
\label{eq:2point_para}
\end{eqnarray}
where $N_{\tau} = 32$ is the extension of the lattice in the time direction. 
We include the contribution of the ground state $(n=0)$ with energy 
$E^0_{\bm{p}}$ and of a first excited state $(n=1)$ with energy $E^1_{\bm{p}}$. 
As discussed in connection with Eq.~\ref{eq:2p}, the parameter $R$ 
indicates the quark-antiquark distance at the sink, which will be 
chosen to enhance the contribution from the pion ground state. 
For this we use the approach as originally proposed in~\cite{Lacock:1995qx}. 
The fuzzed gluon links at the pion sink are created with a 
link/staple mixing of 2 and a fuzzing level of 4. 
The $Z^n_R$ denote the matrix elements,
\begin{equation}
Z^n_R(\bm{p}) \equiv |\left<\Omega | \phi_R | n, \bm{p} \right>|^2 \;,
\label{eq:Zdef}
\end{equation}
which also will yield the `Bethe-Salpeter' amplitudes from which information 
about the structure of the pion can be extracted.

The data for the 2-point function which correspond to the same 
absolute value of the spatial momentum are averaged per configuration. 
The different configurations are then combined in jackknife averages 
with a block-size ranging from 1 to 5. No significant changes in 
the errors of the averages were observed for increasing block-size, 
indicating that there are no significant correlations in our ensemble. 

\subsection{The fuzzing distance $R$}
\begin{figure}
\subfigure 
{
\label{fig:effmassK13230}
\includegraphics[height=85mm,angle=270]{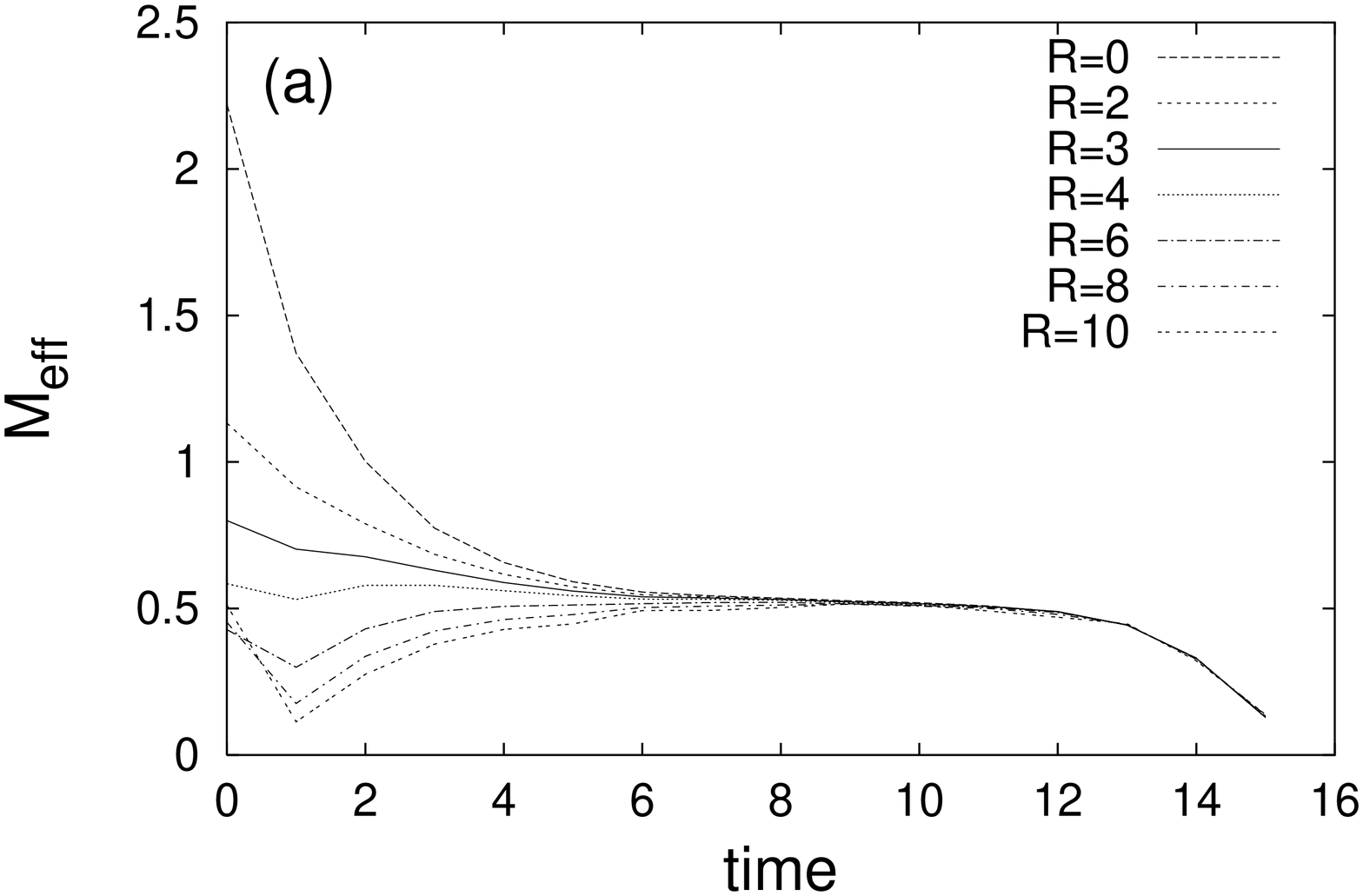}
}
\hspace{0mm}
\subfigure 
{
\label{fig:effenerK13430}
\includegraphics[height=85mm,angle=270]{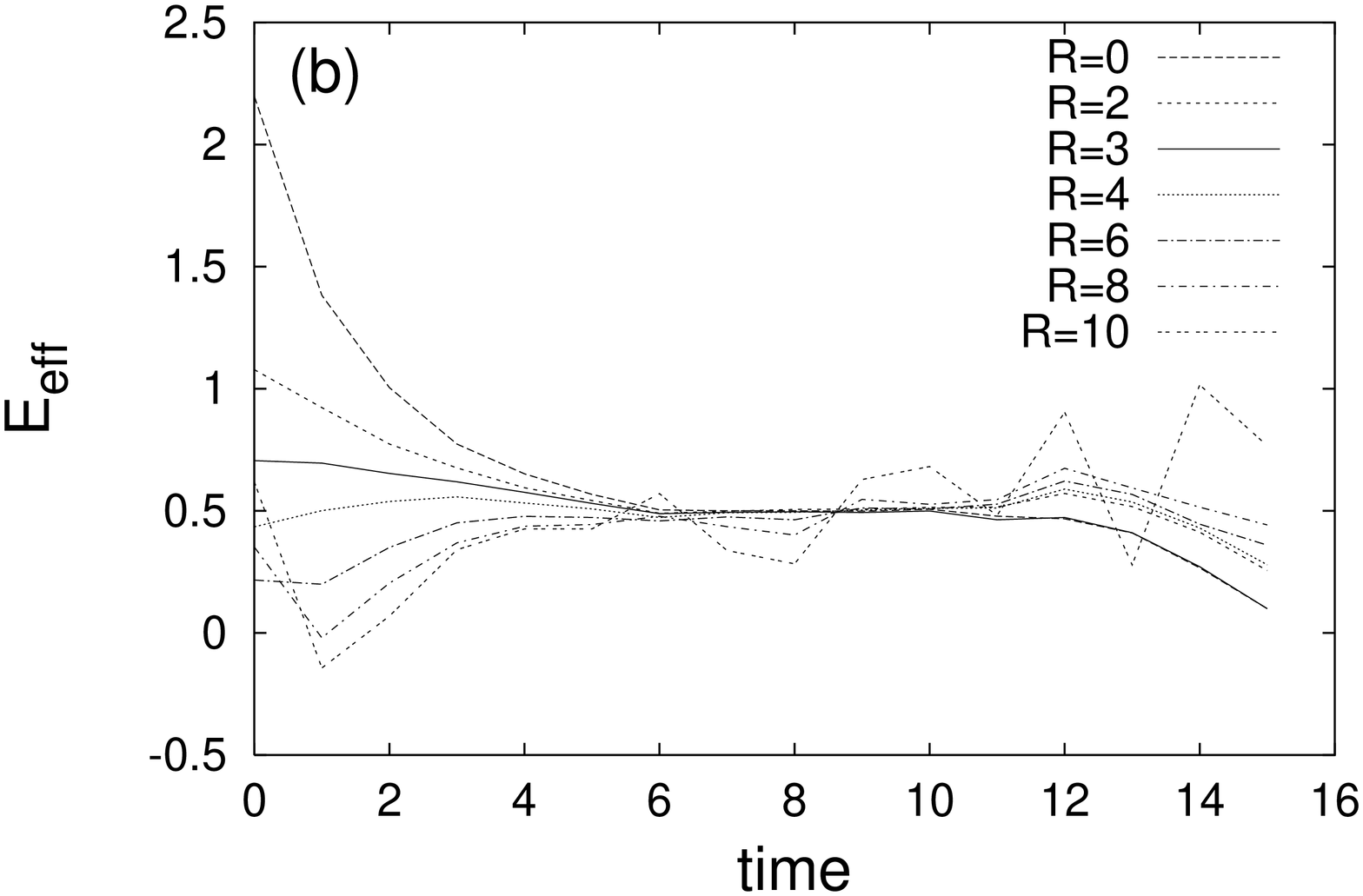}
}
\caption{Effective masses (a) and effective energies (b) for various R.
In (a) the pion mass is $m_{\pi}=970$ MeV, in (b)
$m_{\pi}=540$ MeV and $p^2=0.48$ GeV$^2$.
} 
\end{figure}

To determine the optimal value for the inter-quark distance $R$
we used the jackknife averages to calculate the effective energy of the pion,
\begin{equation}
E_{eff}(t,|\bm{p}|) = \ln(\left<G_{2,R}(t,\bm{p})\right>/
\left<G_{2,R}(t+1,\bm{p})\right>)
\label{eq:effen}
\end{equation}
We varied  $R$ from $0$ to $10$, plotted the effective energy for 
these different fuzzing levels and looked which one stabilizes first. 
An example for the dependence on $R$ is shown in 
Fig.~\ref{fig:effmassK13230} for the effective mass, 
$M_{eff}(t)=E_{eff}(t, 0)$, and for the effective energy in 
Fig.~\ref{fig:effenerK13430} with $|\bm{p}|^2 = 0.48$ GeV$^2$, 
corresponding to the momentum of the pion in our form factor 
extraction below. 
The optimal R-value is somewhat dependent on the pion's momentum
and mass.
After several such tests, we chose $R = 3$ as 
the common extension parameter for all calculations.
Enhancing the ground state contribution
is particularly important for 
the 3-point function, where the distance between 
source and sink is typically small. 

\subsection{Pion masses and energies}
\begin{figure}
\includegraphics[height=85mm,angle=270]{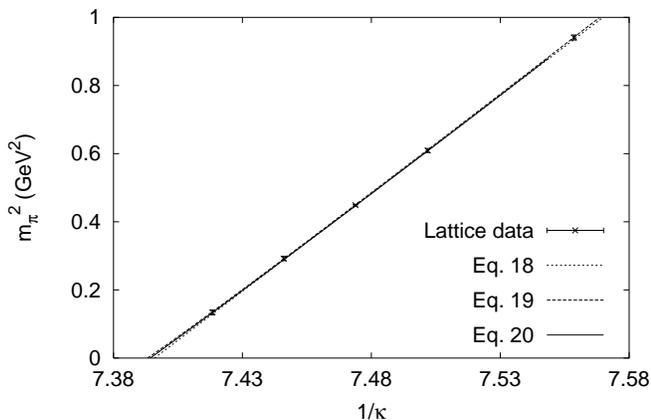}
\caption{\label{fig:pionex} Pion masses as  function of $\kappa$. Lines: 
 extrapolations as indicated.}
\end{figure}

Having chosen the fuzzing distance $R$, we extract the pion 
masses and energies by fitting the jackknife averages to 
Eq.~\ref{eq:2point_para} for both a single state and two states. 
The fit range, the number of included time slices centered around the 
midpoint of our time-grid at $t =16$, is reduced until the 
minimum $\chi^2$ is found and consistency between both fits 
can be checked. In extracting the masses from the $\bm{p} = 0$ 
averages, we found that for a single-state fit a fit range of 
about 15-17 time slices gives
the lowest $\chi^2$. In case of a two-state fit, 
fitting the complete $t$-range (31 time slices) yields the smallest 
statistical error; $\chi^2/dof$ is about the same for different 
fit ranges. We found consistency between the two fits. 
The resulting masses are given in Table~\ref{tab:results2} and 
plotted in Fig.~\ref{fig:pionex} as a function of the inverse 
\begin{table}
\caption{\label{tab:results2} Masses and $\left<r^2\right>_{BS}$ for different $\kappa$-values.}
\begin{ruledtabular}
\begin{tabular}{crcrc}
$\kappa$ 
      & \multicolumn{1}{c}{$m_q$} 
      & $m_{\pi}$ 
      & \multicolumn{1}{c}{$m_\rho$} 
      & $\left<r^2\right>_{BS}$ \\
$0.13230$ & $154$ MeV& $970(4)$ MeV & $1188(6)$ MeV & 0.1414(2) fm$^2$\\
$0.13330$ & $101$ MeV& $780(4)$ MeV & $1053(8)$ MeV & 0.1480(2) fm$^2$\\
$0.13380$ & $75$ MeV& $670(4)$ MeV & $989(9)$ MeV   & 0.1508(2) fm$^2$\\
$0.13430$ & $45$ MeV& $540(6)$ MeV & \multicolumn{1}{c}{--} & 0.1526(2) fm$^2$\\
$0.13480$ & $23$ MeV& $360(9)$ MeV & \multicolumn{1}{c}{--} & 0.1528(4) fm$^2$\\
\end{tabular}
\end{ruledtabular}
\end{table}
of the hopping parameter $\kappa$. They agree very well with the 
results obtained in \textit{e.g.}~\cite{Gockeler:1998fn,Bowler:1999ae}
who use the same action as we do. 
Also shown are extrapolations to the chiral limit, based on the 
following fit functions,
\begin{eqnarray}
m_{\pi}^2 &=& c_1 \left(\frac{1}{\kappa}-\frac{1}{\kappa_{c}}\right)\\
m_{\pi}^2 &=& c_2
\left(\frac{1}{\kappa}-\frac{1}{\kappa_{c}}\right)^{\frac{1}{1+\delta}}\text{ and }\\
\label{eq:fitsharpe}
\frac{1}{\kappa}&=&\frac{1}{\kappa_{c}}+b_1 m_{\pi}^2 +b_2 m_{\pi}^3 \; ,
\label{eq:phen_fit}
\end{eqnarray}
resulting in a slightly different value for $\kappa_{c}$. Quenched 
chiral perturbation theory predicts the second form with $\delta$ 
small and positive \cite{Sharpe:1993gr}. As in \cite{Gockeler:1998fn}, 
we obtain a negative value for $\delta$. Eq.~\ref{eq:phen_fit} is 
a phenomenological fit~\cite{Gockeler:1998fn}. There is not a significant 
difference in the fit quality, so we cannot prefer one fit over 
the other. Therefore, we simply average the different values, yielding 
$\kappa_{c} = 0.13524(4)$, which agrees quite well with values 
obtained from the literature,
$\kappa_{c} = 0.13531(1)$ \cite{Gockeler:1998fn} and 
$\kappa_{c} = 0.13525$ \cite{Bowler:1999ae}. 
The bare quark masses we obtain with 
our $\kappa_{c}$ are given in Table~\ref{tab:results2}. 
They will be used in the (small) mass dependent correction of 
the improved current.

\begin{figure}
\includegraphics[height=85mm,angle=270]{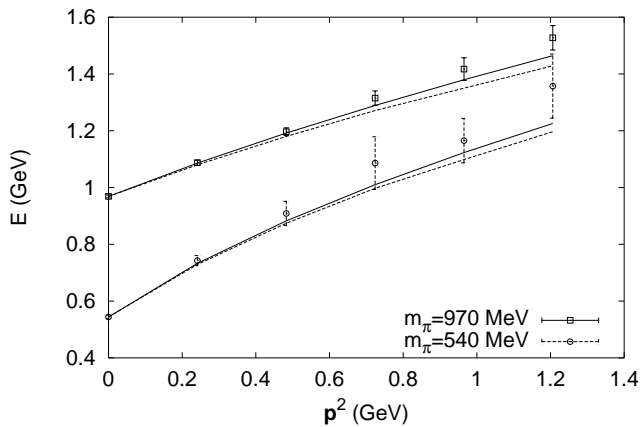}
\caption{\label{fig:disprel} Energy-momentum relation for 
 $\kappa=0.13230$ and $\kappa=0.13430$; solid line: continuum relation, 
dashed: lattice dispersion relation.}
\end{figure}

Proceeding analogously, we have also extracted the pion 
energies $E^0_{\bm{p}}$ for several non-vanishing three-momenta, 
again using single and two-state fits in combination.  
The results for the ground state energy are shown in 
Fig.~\ref{fig:disprel}, together with the prediction from 
the continuum dispersion relation,
\begin{equation}
E_{cont} = \sqrt{m_{\pi}^2 + \bm{p}^2} \; , 
\end{equation}
and from the generally favored lattice dispersion relation,
\begin{equation}
\sinh ^2 \frac{E}{2} = \sinh ^2 \frac{M}{2} + \sum_{i=1}^3 {\sin ^2
 \frac{p_i}{2}} \; . 
\end{equation}
The data deviate from both predictions at higher momenta. However, 
the figure clearly demonstrates that for the momentum relevant 
for our form factor extraction, $\bm{p}^2 = 0.48$ GeV$^2$, 
we are dealing essentially with continuum kinematics.

\subsection{Results for the $\rho$-mass}
In the discussion of the form factor, we will often refer 
to the vector meson dominance model. For completeness and 
as a further test for our methods, we have also extracted 
the mass of the lowest vector meson, the $\rho$-meson. 
Proceeding analogously as for the pion, Eq.~\ref{fig:2p}, we 
consider a 2-point function with source and sink operators of the form
\begin{equation}
V_{i} =  {\bar \psi}\; \gamma_{i}\; \psi \; ,
\end{equation}
which project onto the polarization state $i$ of a vector meson. 
With the same boundary conditions as for the pion, the 
2-point function for three-momentum $\bm{p}=0$ was then 
fitted to a $\cosh$-form as in Eq.~\ref{eq:2point_para}. 
We averaged the polarization states $i = 1, 2, 3$. 
The results for the three highest parameters $\kappa$ are 
shown in Table~\ref{tab:results2}. They agree with the values 
obtained in ~\cite{Gockeler:1998fn,Bowler:1999ae}. 
For the remaining two $\kappa$ values our statistics were 
not sufficient. When needed later for the comparison with 
the vector meson dominance model for the form factor, we 
will for simplicity use the values from~\cite{Gockeler:1998fn}.

\subsection{Bethe-Salpeter amplitude and $\left<r^2\right>_{BS}$}
\label{sec:BSA_RMS}
To obtain the `Bethe-Salpeter (BS) amplitudes' 
or 'wave functions' $\Phi (R)$, we use the $Z$ factors 
extracted from a fit of the 2-point function 
(Eq.~\ref{eq:2point_para}), since for a pion at rest, 
\begin{equation}
\Phi(R)=\sqrt {Z^0_R (\bm{0}) \, / \, Z^0_0 (\bm{0})}.
\label{eq:BSA}
\end{equation}
We simultaneously fitted the jackknife averages for $R$ ranging 
from $0$ to $10$. Again the results from both the 
single- and a two-state parametrization were used to check the consistency. 

\begin{figure}
\includegraphics[height=85mm,angle=270]{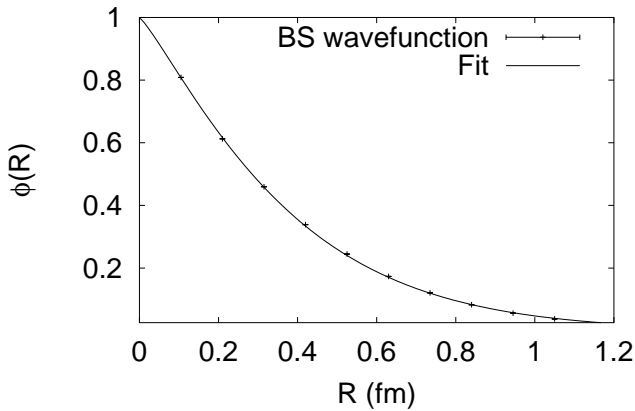}
\caption{\label{fig:BS_WF} BS wave function for $\kappa=\mathit{0.13230}$.}
\end{figure}

The same results for the wave function, but with smaller 
errors, were obtained from a fit to the plateau of the ratio
\begin{equation}
{\tilde \Phi}(R) = \frac{G_{2,R} (t, \bm{0})}{G_{2,0}(t, \bm{0})} \; .
\end{equation}
An example for a BS-wave function is shown in 
Fig.~\ref{fig:BS_WF} for the heaviest pion; only the result 
of the second method is displayed. 
The wave function can be used to obtain an estimate of 
the mean square charge radius of the pion according to
\cite{Hecht:1992uq,Laermann:2001vg}
\begin{equation}
\left<r^2\right>_{BS}:=\frac{1}{w} \; \frac{\int
 d^3\vec{r}\;\vec{r}^{\,2}\;\Phi^2(|\vec{r}|)}{\int
 d^3\vec{r}\;\Phi^2(|\vec{r|})}\;.
\end{equation}
The factor $w$ is included to reflect the uncertainty in 
the resulting $\langle r^2 \rangle$. If one assumes that quark and 
antiquark are always located on opposite sides of the 
meson center of mass at distance $r/2$, one has $w = 4$; 
assuming that the quarks move uncorrelated around the 
center-of-mass, 
$w  = 2$ should be used. The BS-radii as a function of 
the pion mass are given in Table~\ref{tab:results2} for $w = 2$. 
The values are much lower than the physical value 
of 0.439(8) fm$^2$ \cite{Amendolia:1986wj}. Moreover, the mean square radius 
is seen to be almost independent of the pion mass
in the investigated range. 
We will comment on this in more detail in Section~\ref{sec:dis}.

\section{The 3-point function}
\label{sec:3point}
For the 3-point function, already introduced 
in Ch. 2, we now consider a pseudo-scalar source 
at $t = 0$, a sink at $t_f = 11$ and a coupling to the 
photon at $t$ with $0 < t < t_f$. 
Barad \textit{et al.}~\cite{Barad:1984px} pointed out 
that current conservation provides an important 
numerical test which relates the 2- and 3-point 
functions. Translated into momentum space and for 
our periodic boundary conditions, this relation reads
\begin{equation}
G_3(t_f, t; \bm{p}, \bm{p}) - G_3(t_f, t'; 
\bm{p}, \bm{p}) = G_2 (t_f,\bm{p}) \; ,
\label{eq:chargecons}
\end{equation}
where, in the second term on the left hand side, 
$ t_f < t' < N_{\tau}$. It amounts to considering the 
total charge that leaves the source in the forward 
and backward direction in time and guarantees that 
$F(Q^2) = 1$ at $Q^2 = 0$. 
This relation holds for each background gauge field configuration separately 
and thus also for configuration averages. We have verified 
that our results for the conserved current
satisfy this relation 
to an accuracy better than ${\cal O}(10^{-5})$.

In order to obtain $F(0)$ for the
renormalized local and the improved current,
we again exploited Eq.~\ref{eq:chargecons}.
The lhs of this equation was averaged over pairs of values $t$ and $t'$
symmetric around $t_f$ and normalized by the 2-point function.
Utilizing the $Z_V$ factor
from~\cite{Bhattacharya:2000pn} gives $F^I(0) = F^{RL}(0) = 1$ within
a jackknife error of 1 \%.
Alternatively, we could have applied this method to
independently extract $Z_V$ as done {\it e.g.} in~\cite{Bakeyev:2003ff}.
However, for consistency we used the entire set
of improvement parameters from~\cite{Bhattacharya:2000pn}.

While the above method allows one to determine $F(0)$, 
we now describe how we extract the form factor for\\
$Q^2 > 0$. 
As in the 2-point function we allow two states to 
contribute and parametrize the 3-point function, Eq.~\ref{eq:3p}, as
\begin{eqnarray}
G_{3,\mu}(t_f,t;\bm{p}_f,\bm{ p}_i) = 
\sum_{m=0}^{1}\sum_{n=0}^{1}
\sqrt {Z_R^m (\bm{p}_f) Z_0^n (\bm{p}_i)}\nonumber \\
\times\;\;\left<m,\bm{p}_f| j_\mu (0) |n,\bm{p}_i \right>
e^{-\, E^m_{\bm{p}_f} \, (t_f - t)  -\, E^n_{\bm{p}_i} \, t}\;,
\end{eqnarray}\\
where $(m,n)\neq(1,1)$.
Contributions from, for example, the production of pion pairs, as well
as 'wrap around effects' due to the propagation of states beyond $t_f$
are exponentially suppressed ($ < \mathcal{O}$$(e^{-5})$); similarly,
an elastic contribution from the excited state was estimated to be of
the order of $1\%$ or less.  All these effects are not reflected in
the chosen parametrization. The inelastic transitions $0
\leftrightarrow 1$ are included to better describe the data. However,
it should be understood that the state labelled $1$ parametrizes
contributions from all possible excited states.  Therefore we do not
interpret our parameters as the energy or the transition form factors
corresponding to a single genuine excited state.
 
Since for a given momentum the pion is the state with the 
lowest energy, one gets with Eqs.~\ref{eq:lorentz_dec} 
and~\ref{eq:normalisation} 
\begin{widetext}
\begin{eqnarray}
G_{3,\mu}(t_f,t;\bm{p}_f,\bm{ p}_i) = F(Q^2)\;
\frac{(p_f+p_i)_{\mu}}{2\sqrt{E^0_{{\bf p}_f} E^1_{{\bf p}_i}}}\sqrt {Z_R^0 
(\bm{p}_f)Z_0^0 (\bm{p}_i) }\; e^{- E^0_{\bm{p}_f }\; (t_f - t)\,  - E^0_{\bm{
p}_i}\; t }~~~~~~~~~~~~~~~~~~~~~~ \nonumber\\
+\left\{\sqrt {Z_R^1 (\bm{p}_f) Z_0^0 (\bm{p}_i)}\left<1,\bm{p}_f| 
j_\mu (0) |0,\bm{p}_i \right>e^{-\, E^1_{\bm{p}_f} \, (t_f - t)  -\, 
E^0_{\bm{p}_i} \, t} \;+ (1 \leftrightarrow 0) \right\}\;.
\label{eq:param}
\end{eqnarray}
\end{widetext}
In the simulations, we took $|\bm{p}_f|=|\bm{p}_i|=\sqrt{2}|\bm{p}_{min}|$, where
\begin{equation}
|\bm{p}_{min}| = \frac{2 \pi}{N_{\sigma} a}
\end{equation}
is the minimal momentum for a lattice with $N_{\sigma}$ lattice 
points in the spatial extension. In our case, it amounts to 
\begin{equation}
\bm{p}_i^2 = \bm{p}_f^2 = 0.48 \text{ GeV}^2 \; .
\end{equation}
For our analysis we use the fourth component of the current, $\mu = 4$. 
With our choice of momenta the kinematical factors in 
the fit function, Eq.~\ref{eq:param}, therefore simplify considerably,
see Eq.~\ref{eq:prefac}; note also that the $t$-dependence of the first term in Eq.~\ref{eq:param}
vanishes.
As a result, the form factor is more easily extracted without 
restricting the simulation too much. Different momentum 
transfers are obtained by varying the relative orientation 
of $\bm{p}_i$ and $\bm{p}_f$.

\subsection{Extraction of parameters}
\label{Ext_par}
\begin{figure}
\subfigure 
{
\includegraphics[height=85mm,angle=270]{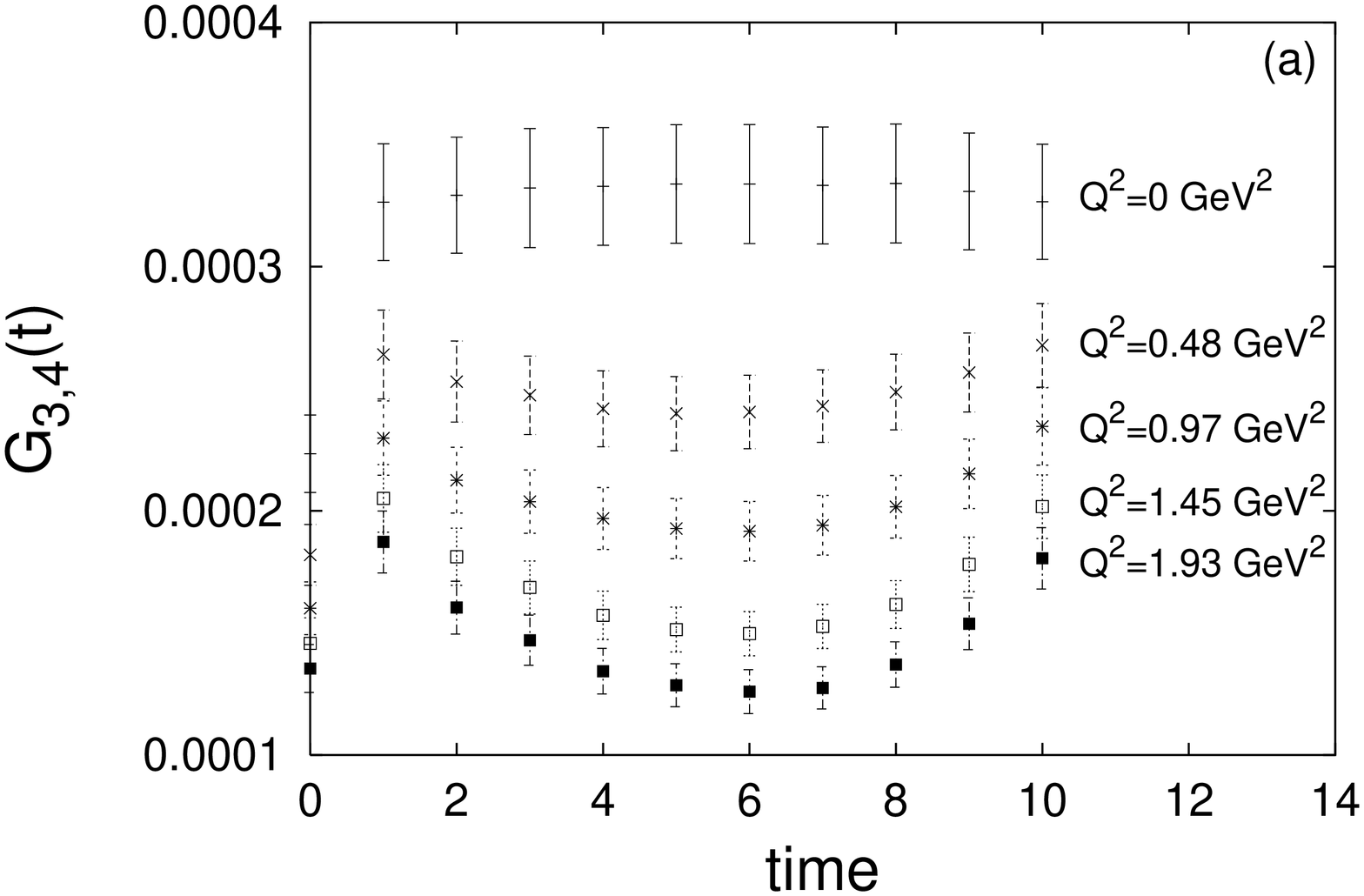}
\label{fig:3point_all_Q2}
}
\subfigure 
{
\includegraphics[height=85mm,angle=270]{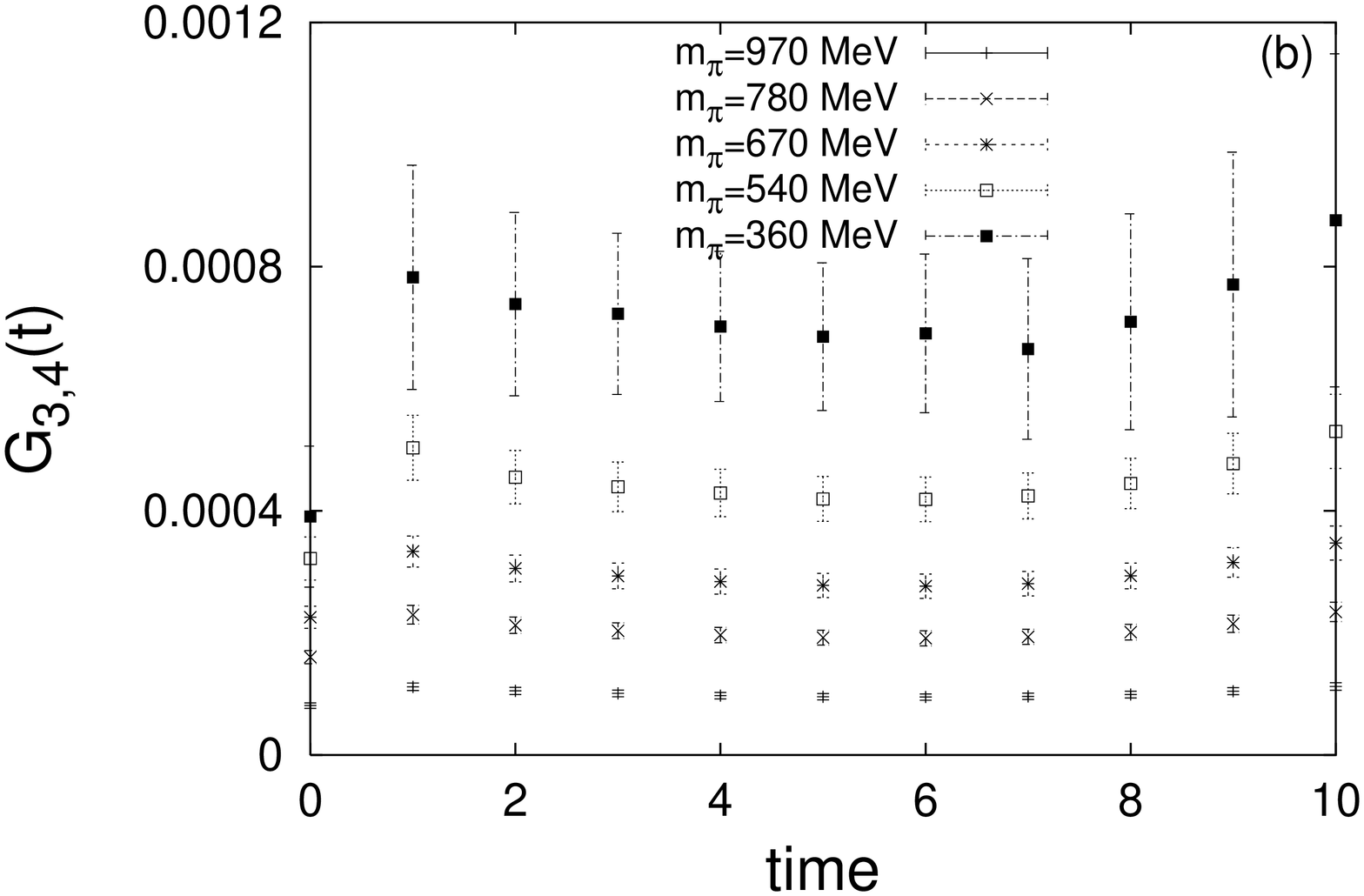}
\label{fig:3point_all_k}
}
\caption{Improved 3-point function for (a) different $Q^2$ at 
    $m_{\pi}=780$ MeV and (b) different $m_{\pi}$ 
at $Q^2=0.97$ GeV$^2$.}
\end{figure}

We begin by averaging the 3-point correlation 
functions which have the same four-mo\-men\-tum transfer 
squared and then again combine the configurations in 
jackknife averages. Typical jackknife averages of the 
3-point function are shown in Fig.~\ref{fig:3point_all_Q2} 
for the one but highest pion mass for various momentum 
transfers and for different masses at fixed momentum 
transfer in Fig.~\ref{fig:3point_all_k}. If only the 
pion ground state would contribute, there would be 
no $t$-dependence in this quantity. The data however 
clearly indicate the admixture of an excited state. 
This is especially seen at high $Q^2$ and for low pion 
masses, where there is no time-slice where it is safe to 
assume that only the pion ground state is present.

We therefore chose to proceed by simultaneously fitting
the parameters in the 2-point and 3-point function,
Eq.~\ref{eq:2point_para} and Eq.~\ref{eq:param}, respectively. 
We hereby exploit the fact
that certain parameters appear in both Green's functions.
In the case of the 2-point function,
we fitted the energies, $E^0$ and $E^1$ and the $Z$ factors
over the complete time interval, $1 \leq t \leq N_\tau-1$.
In the 3-point function, we fit the same energies and $Z$-factors,
and in addition the form factor $F(Q^2)$ and the transition
matrix elements over the interval $t_i < t < t_f$.
We assume that the energies and $Z$ factors only depend on the
magnitude of the three-momenta and use the fact that we
chose $|{\bf p}_f |=|{\bf p}_i |$.
The fits are done for each value of $Q^2$ separately. The values for
$\chi^2/dof$ lie between 0.15 and 0.40, depending on mass and momentum
transfer. The energies and $Z$ factors we obtain from our fits at
different $Q^2$ agree to high accuracy because they are largely
determined by the 2-point function.

To compare with earlier work 
\cite{Martinelli:1988bh,Draper:1989bp}
we also extracted an estimate for the form factor
from the ratio of 3- and 2-point functions.
However, the assumption of just a single state contributing
is at the basis of this method. Correspondingly,
we found differences ranging from $5 - 10 \%$ for
$F(Q^2)$ between the ratio method and our combined fit
method, where inclusion of an excited state clearly
improved the fit quality. The size of the difference
depends on the pion mass and the momentum transfer, which 
influences the flatness of the 3-point function in the middle 
between source and sink. All our results in the next 
section are therefore based on the fit method.

\section{\label{sec:dis} Results and Discussion}

We will now discuss the form factors obtained 
from the procedure described in the previous Chapter. 

\subsection{Dependence on the current}
We first compare the form factors we obtain with the 
local current, Eq.~\ref{eq:lc}, the 
conserved current, Eq.~\ref{eq:cc}, and the improved current, 
Eq.~\ref{eq:ic}. Only the improved current ensures that 
there are no corrections to ${\cal O}(a)$ with our action. 

\begin{figure}
\subfigure 
{
\label{fig:cc_K13230}
\includegraphics[height=85mm,angle=270]{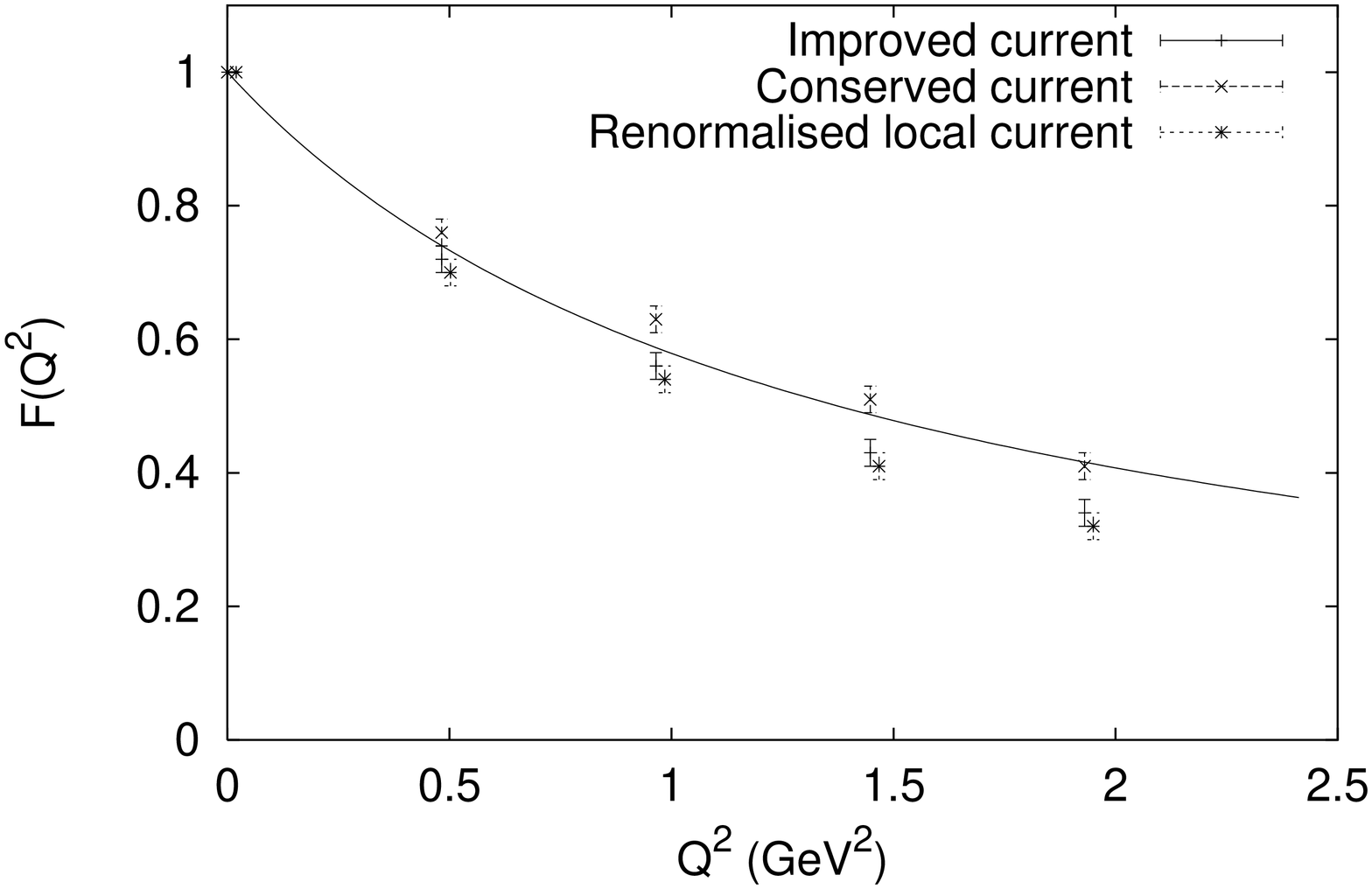}
}
\subfigure 
{
\label{fig:cc_K13480}
\includegraphics[height=85mm,angle=270]{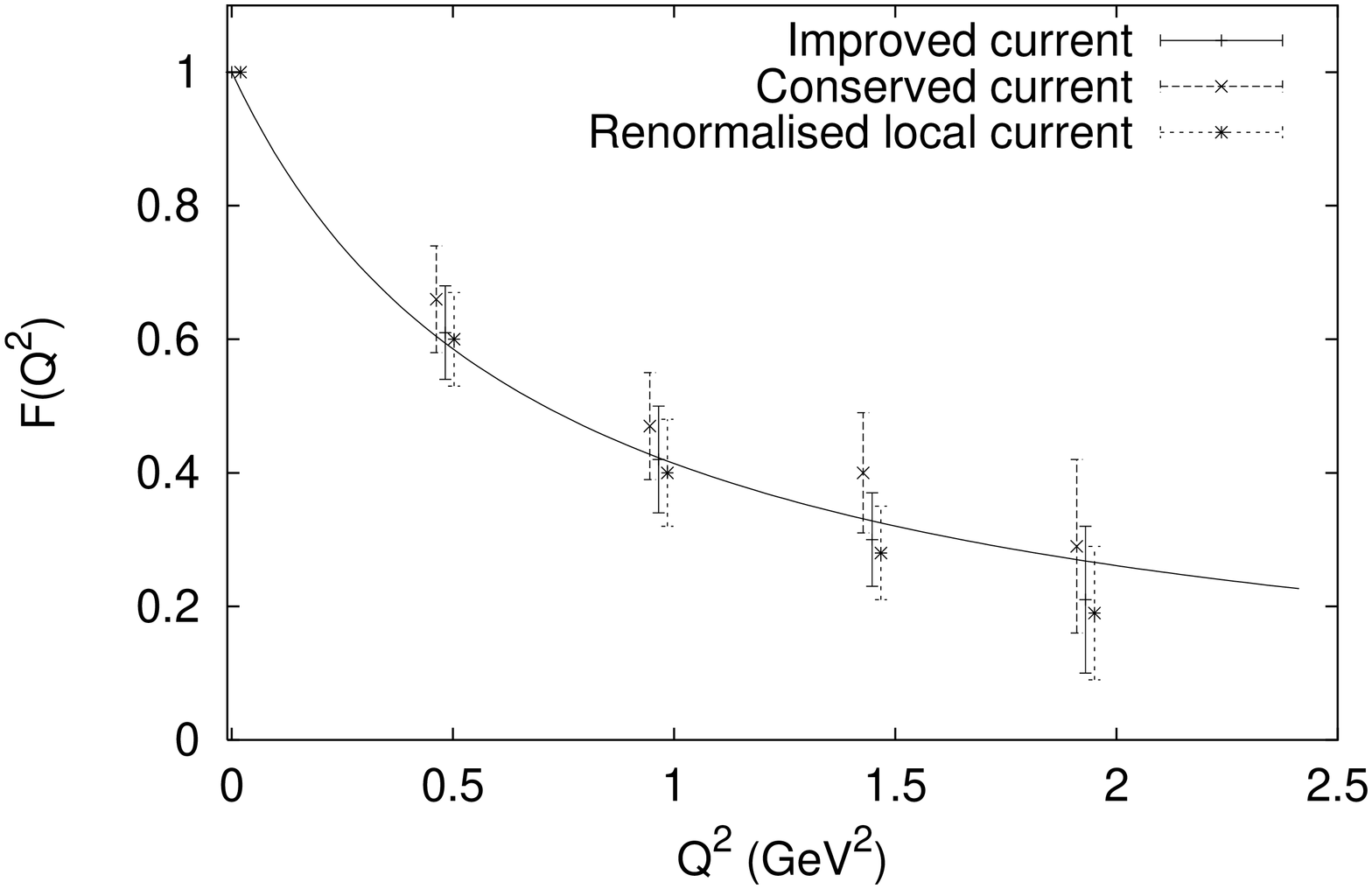}
}
\caption{Form factors extracted from different currents as a function 
of $Q^2$ for 2 pion masses: $m_{\pi}=970$ MeV (top)
and $m_{\pi}=360$ MeV (bottom). Solid curve: 
VMD model prediction with $m_V = m_{\rho}$ 
taken from \protect\cite{Gockeler:1998fn}. 
Data shifted horizontally for clarity.}
\end{figure}

The form factors for $Q^2 > 0$, obtained through the simultaneous fit
procedure, are shown in Figs.~\ref{fig:cc_K13230} and~\ref{fig:cc_K13480}
for two different masses.
As can be expected from Fig.~\ref{fig:3point_all_Q2}, the same fit procedure yields
values for $F(0)$ with an error comparable to the results for
low $Q^2$. However, we use the method discussed in connection with
Eq.~\ref{eq:chargecons}, to extract $F(0)=1$ to higher accuracy.

We first of all observe that, as 
expected, the results for our heaviest pion, 
$m_\pi = 970$ MeV, are much more accurate 
than for the lightest pion, $m_\pi = 360$ MeV. 
However, even with the larger error bars, a rather 
smooth $Q^2$-dependence is seen also in the latter case.
We further observe that the differences between conserved and improved current 
grow with momentum transfer
and decreasing mass, resulting, 
in particular for the light pion mass, in a 
substantial correction. 
The differences between the improved and the renormalized local
current are due to the tensor term.
Although this contribution increases with $Q^2$
the improved form factor stays very close to 
the result for the renormalized local current, 
also for the light pion and out to our largest 
momentum transfer. 
That these two form 
factors are so close is due to the fact that the 
contribution of the tensor term in the improved 
current is small. Since the matrix element 
of the tensor operator can become comparable in 
size (up to 70 \%) to the local current operator, this smallness 
is due to the fact that the coefficient, $c_V = -0.107$, 
determined in~\cite{Bhattacharya:2000pn} is rather 
small; the preliminary value obtained by the ALPHA 
collaboration~\cite{Guagnelli:1998db} is much 
larger, $c_V = -0.32$. 
Since the improved current is a linear combination 
of the local and the tensor term, Eq.~\ref{eq:renorm}, the change in 
the improved current 
due to a change in $c_V$ is straightforward. The 
difference between both values of $c_V$ is of order $a$, 
resulting in improved currents which are different only at order $a^2$.
This shows that
${\cal O}(a^2)$ effects can still become as large as 10 \% at higher
$Q^2$ values and low masses.

\subsection{The form factor and vector meson dominance}
\begin{figure}
\includegraphics[height=85mm,angle=270]{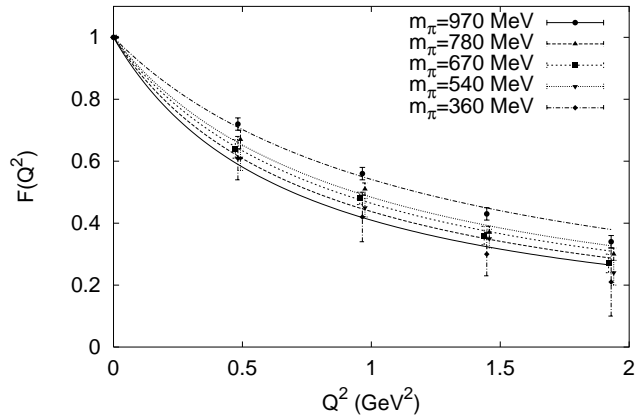}
\caption{\label{fig:ff_vmd_plot} Improved form factor as a function of $Q^2$ for 
different $m_\pi$. Lines: fits to VMD-form, Eq.~\ref{eq:vmd}.}
\end{figure}

As was already observed in~\cite{Draper:1989bp}, the 
lattice results for the form factor show a behavior 
expected from vector meson dominance. 
In Figs.~\ref{fig:cc_K13230} and~\ref{fig:cc_K13480}, 
we show the prediction for the form factor if we 
use the simple monopole form, Eq.~\ref{eq:vmd}, 
with $m_V = m_\rho$, the lattice $\rho$-mass at the 
same $\kappa$-value \cite{Gockeler:1998fn}.
At large pion mass the VMD-prediction describes
the form factor based on the conserved current rather well
but lies substantially above the results from the improved
current.
However, at lower pion masses the model prediction shifts
toward the improved form factor results.  
To investigate this point in more detail, 
we fitted the improved form factors, using the vector meson 
mass $m_V$ as a fit parameter, 
omitting the point at the highest $Q^2$ value.
The parametrization works well and results are shown 
in Fig.~\ref{fig:ff_vmd_plot}. 
Table~\ref{tab:results3}
 \begin{table}
\caption{\label{tab:results3} $m_{\rho}$ from the 2-point function 
\protect\cite{Gockeler:1998fn}, fitted $m_V$ and $\left<r^2\right>$ 
for different $m_\pi$ values.}
\begin{ruledtabular}
\begin{tabular}{crrr}
$m_{\pi}$ 
    & \multicolumn{1}{c}{$m_\rho$} 
    & \multicolumn{1}{c}{$m_V$} 
    & \multicolumn{1}{c}{$\left<r^2\right>$} \\
$970(4)$ MeV & $1169(3)$ MeV  & $1086(26)$ MeV & $0.197(9)$ fm$^2$\\
$780(4)$ MeV & $1032(4)$ MeV & $968(26)$ MeV & $0.249(13)$ fm$^2$\\
$670(4)$ MeV & $966(6)$ MeV & $931(26)$ MeV & $0.269(15)$ fm$^2$\\
$540(6)$ MeV & $901(6)$ MeV & $882(36)$ MeV & $0.299(24)$ fm$^2$\\
$360(9)$ MeV & $841(24)$ MeV & $833(75)$ MeV & $0.34(6)$ fm$^2$\\
\end{tabular}
\end{ruledtabular}
\end{table}
compares the fitted $m_V$ values to the $\rho$-mass 
extracted from 2-point functions~\cite{Gockeler:1998fn}.
The two values can be seen 
to come closer together as the pion becomes lighter, 
suggesting a better agreement of the improved results 
with the simple vector 
meson dominance model for lower pion masses. 
However, as we will see later, in the physical limit, using
$m_V = m_\rho$ fails to describe the experimental data accurately.

\subsection{Determination of the charge radius}
It is well known that the slope of the form factor 
is related to the mean-square charge radius of the pion,
\begin{equation}
\left. \frac{\partial F(Q^2)}{\partial Q^2}
\right|_{Q^2=0}=\frac{1}{6}\left<r^2\right>\; .
\label{eq:rms_ff}
\end{equation}
In contrast to the charge-radius extracted from 
the Bethe-Salpeter amplitude, this determination 
of $\left<r^2\right>$ is not based on any 
specific assumptions about the quark motion 
inside the pion. As the vector meson dominance 
model works very well for low $Q^2$, we show 
for simplicity the mean square charge radius of the 
pion obtained from the monopole fit, which yields
\begin{equation}
\left< r^2 \right>=\frac{6}{m_V^2} \; .
\end{equation}
In the following we only work with the improved current.
By looking at the values in Table \ref{tab:results3} 
and in Fig.~\ref{fig:rms_comp_ex}, we observe that 
the $\left<r^2\right>$ extracted from the form factor 
shows a considerable mass dependence. This is in 
contrast to the BS results, which are also shown. 
Moreover, these results, which we obtained in 
Section~\ref{sec:BSA_RMS}, are considerably lower 
than the value we extract from the form factor. 
As already discussed by Gupta \textit{et al.}~\cite{Gupta:1993vp}, 
this can be due to how one treats the center of mass of 
the two quarks. However, as these 
authors also point out, the value extracted 
through the form factor contains contributions 
that are not included when calculating 
$\left<r^2\right>$ from the Bethe-Salpeter 
amplitude.

The $\left<r^2\right>$ values obtained from the 
form factor can be seen to get closer to the 
physical value of $\left<r^2\right> = 0.439(8)$ fm$^2$
as the pion mass decreases. 
This led us to try several extrapolations to the 
physical limit which will be described in the next section.

In addition to the two methods discussed above, 
there is another method to obtain the charge 
radius of the pion from lattice QCD. This method is based 
on calculating density-density correlations or 4-point functions 
for the pion~\cite{Barad:1984px,Wilcox:1986ge}. 
It has recently been used by 
Alexandrou \textit{et al.}~\cite{Alexandrou:2002nn}
for densities at equal times. 
However, there are
difficulties in the extraction of $\left<r^2\right>$ from density-density
correlations as discussed in detail by 
Burkardt \textit{et al.}~\cite{Burkardt:1995pw} and Wilcox~\cite{Wilcox:2002zt}.

\subsection{Extrapolation in $m_\pi$}
To obtain more physical results, one can try to 
extrapolate in the pion mass. We take 
$\left<r^2\right>$ as the quantity to extrapolate, 
since it is known experimentally and its extrapolation 
has been discussed in the literature. We consider 
three different types of extrapolations. 
From chiral perturbation theory ($\chi PT$), one knows 
the 1-loop result~\cite{Gasser:1984yg},
\begin{equation}
\left< r^2 \right>^{one-loop}_{\chi PT}= c_1+c_2 \ln m_{\pi}^2\;.
\end{equation}
In our fit, we will treat $c_1$ and $c_2$ as free parameters.
In \textit{quenched} $\chi PT$, the radius is constant 
at this order of expansion~\cite{Colangelo:1998ch}. 
It is however expected that this situation will change 
at the two-loop level, which will introduce terms like
\begin{equation}
\left<r^2\right>^{two-loop}_{q\chi PT} ~ \sim ~ \; d_1 \; 
\frac{1}{m_\pi^2} + d_2 \,\ln {m_\pi^2} \; + d_3 \,m_\pi^2 \; ,
\end{equation}
including a term linear in $m_{\pi}^2$ which,
for our pion masses, can be expected to yield the dominant 
contribution \cite{Colangelo:priv}. 
We therefore only tried a form containing a constant plus
a term linear in $m_{\pi}^2$ 
\begin{figure}
\includegraphics[height=85mm,angle=270]{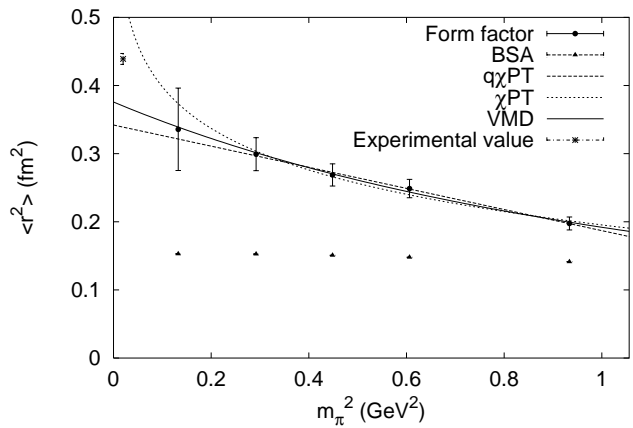}
\caption{\label{fig:rms_comp_ex} Pion charge radii extracted from the 
form factor and extrapolations in $m_\pi^2$. 
The BS-radii are included for comparison.
Experimental result from \protect\cite{Amendolia:1986wj}}
\end{figure}
\begin{figure*}
\includegraphics[height=170mm,angle=270]{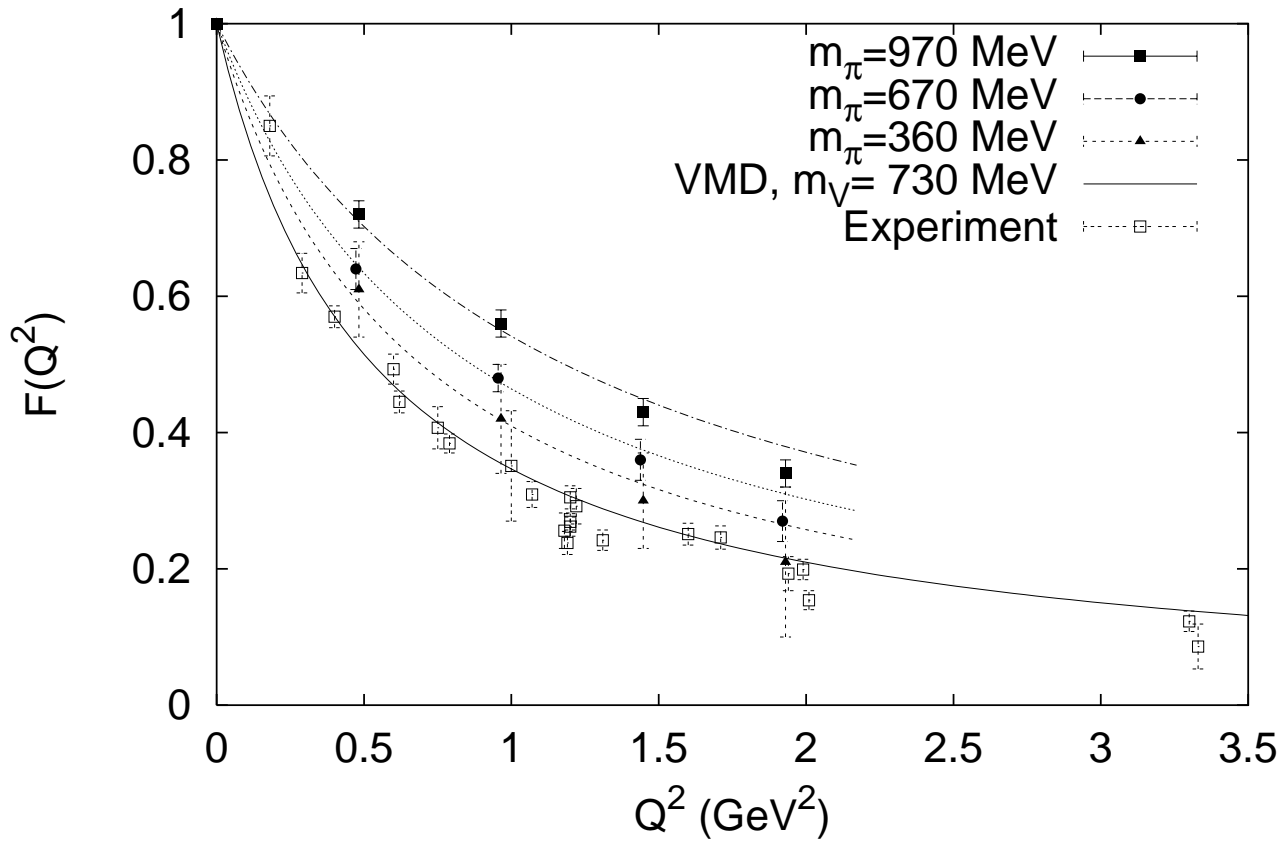}
\caption{\label{fig:ff_exp_comp} Improved form factor for different $m_{\pi}$, compared to experiment
\protect\cite{Bebek:1978pe,Volmer:2000ek}; broken lines as in Fig.~\protect\ref{fig:ff_vmd_plot}.} 
\end{figure*}

We have observed that our form factor data 
can be well described by a monopole form as suggested by
simple vector meson dominance. 
Therefore, we use this model to 
obtain an additional extrapolation. 
Since it can be seen that $m_V$, just like $m_\rho$, 
scales approximately linear with $m_\pi^2$, one 
arrives at,
\begin{equation}
\left< r^2 \right>_{VMD}=\frac{6}{(b_1+b_2 \, m_{\pi}^2)^2} \; .
\end{equation}
The three extrapolations are plotted in 
Fig.~\ref{fig:rms_comp_ex}. 
The extrapolated value 
is seen to depend strongly on the method chosen. 
The VMD ansatz describes the data best.
If we use this ansatz to extrapolate to the physical
pion mass we find 
$\left<r^2\right>=0.37(2)$ fm$^2$, which lies below 
the experimental value of $0.439(8)$ fm$^2$. 
This could 
clearly be due to the extrapolation chosen, but 
also due to assumptions and approximations, 
such as quenching, inherent in our approach.

\subsection{Comparison with experiment}
In Fig.~\ref{fig:ff_exp_comp} we show our results 
together with the available 
measurements~\cite{Bebek:1978pe,Volmer:2000ek}. 
For clarity we only show our results for three $\kappa$-values.
As can be clearly seen, all our calculated form 
factors lie above the measured values. 
Nevertheless, a continuous trend toward
the experimental values can be observed and
we come quite close to them.
Whether only a straightforward further lowering 
of the pion mass will resolve the remaining 
discrepancy between our lattice calculations and 
experiment is not clear.
The solid line in the figure shows the monopole form using
$m_V^2 = 6 / \left< r^2 \right>_{exp}$
with the experimentally measured charge radius
$\left< r^2 \right>_{exp}$.
This describes the experimental data quite well
also away from $Q^2 = 0$. However, the corresponding
vector meson mass at 730 MeV is significantly lower
than the $\rho$-mass, emphasizing 
that
the VMD-inspired monopole description provides a successful
parametrization of the form factor data, but does not hold
in detail.

\section{Summary and Conclusions}

We have presented calculations for the pion form factor that improved 
and extended previous work in several respects. We have pushed the form 
factor calculations for a large range of $Q^2$ towards much lower pion 
masses than before. In doing this, we have worked in a framework that 
guarantees the absence of ${\mathcal O}(a)$ corrections. This meant a 
consistent use of an improved action with the concomitant improved 
conserved current. It was shown that use of this improved current leads 
to significant changes over results based on the conserved Noether 
current for this action, which still contains ${\mathcal O}(a)$ 
corrections at finite $Q^2$. We chose kinematics where the initial and 
final pion momentum had the same absolute value, which leads to practical 
simplifications when extracting the form factor. For the momenta we use, 
we confirmed that energy and momenta are sufficiently close to satisfying 
a continuum dispersion relation.

Our results for the form factor were seen to smoothly vary with pion mass. 
The lower $Q^2$ results could all be described quite well by a simple 
monopole form factor. The 
fitted range parameter $m_V^{-1}$ was, for each $\kappa$-value, found to 
be close to the corresponding lattice $\rho$-mass. The agreement between 
the two values got closer for decreasing pion mass, indicating better 
agreement with the vector meson dominance model.

The form factor can be used to extract the mean square charge radius of 
the pion. The values we obtained show that the estimates for 
$\left<r^2\right>$ based on the Bethe-Salpeter amplitude are qualitative 
as well as quantitative not very reliable. Disagreement of up to a factor 
two was found with the form factor based values, which showed also a quite 
pronounced mass-dependence in contrast to the BS results. Extrapolations 
of our charge radii towards the physical pion mass were shown to lead to 
no unique prediction. The best description of the results at our pion 
masses was provided by a vector meson dominance model. When extrapolated 
to the physical pion mass, it yields a value for $\left<r^2\right>$ 
about 15\% below the experimental value. For an extrapolation inspired by 
(quenched) chiral perturbation theory our pion masses are too high to be 
sufficiently sensitive to the predicted $\ln m_\pi ^2$ terms.

When compared to the experimental form factor, it could be seen that our 
results consistently approach the data from above over the entire range 
of $Q^2$ we consider. While gauge invariance fixes all form factors at 
$Q^2 = 0$ to $F(0) = 1$, we see that the calculated form factor at $Q^2>0$ 
comes close to the experimentally determined shape, and to a monopole 
parametrisation. This is a nice confirmation that lattice QCD indeed 
describes a non-perturbative feature such as a pion form factor quite 
realistically and in detail. However, a straightforward extension of our 
approach to even lower pion masses or higher $Q^2$ is not necessarily the 
way to proceed to close the last gap to the experiment. Improvements of 
our approach and other lattice methods will become necessary. Corrections 
of order ${\mathcal O}(a^2)$, for example, will become increasingly 
important and one also has to understand the role of dynamical quarks, 
which are neglected in the quenched approximation. 

As is well known, Wilson fermions have the major disadvantage that chiral 
symmetry is broken, already at $\mathcal{O}(a)$. This was one of the 
reasons improvement was invented and why we chose a framework where action 
and operators where consistently improved and only corrections to order 
${\mathcal O}(a^2)$ show up. Another method, among others, is the 
introduction of a fifth dimension and use of so-called domain wall 
fermions. Chiral symmetry is then not tied to taking the continuum limit. 
The price one pays is a substantial increase in the computer time. The RBC 
collaboration has chosen this approach and first results can be found 
in~\cite{Nemoto:2003ng}. In this paper pion masses down to 390 MeV are 
used, albeit on a coarser lattice. Their results obtained so far at 
two low $Q^2$ points, based on the renormalised local current, seem to 
agree reasonably well with our values. Differences in the implementation 
of chiral symmetry show up at ${\mathcal O}(a^2)$.

An open question is of course the significance of the quenched 
approximation. Alexandrou \textit{et al.}~\cite{Alexandrou:2002nn} have 
calculated density-density correlations for the $\pi$, $\rho$, $N$ and 
$\Delta$ in quenched as well as unquenched lattice QCD. In contrast to the 
$\rho$ and $\Delta$, only rather small effects are seen for the $\pi$ for 
values of $m_\pi$ around $600$ MeV. The study of effects due to dynamical 
quarks, clearly more important at lower pion masses and high $Q^2$, is an 
area where further work is necessary.

\begin{acknowledgments}
The work of J.v.d.H and J.H.K. is part of the research program of the 
Foundation for Fundamental Research of Matter (FOM) and the National 
Organization for Scientific Research (NWO) of The Netherlands. The research 
of E.L. is partly supported by Deutsche Forschungsgemeinschaft (DFG) 
under grant FOR 339/2-1. The computations were performed at 
the John von Neumann Institute for Computing (NIC), J\"ulich 
and at SARA, Amsterdam under grant SG-119 by the Foundation for National
Computing Facilities (NCF).

The authors thank G. Colangelo for helpful comments concerning the 
extrapolation in the pion mass. J.v.d.H. thanks S. Sharpe for many 
stimulating discussions.
\end{acknowledgments}


\end{document}